# Unraveling Intertwined Orders in the Strongly Correlated Kagome Metal $CsCr_3Sb_5$

Liangyang Liu[1*], Yidian Li[1*], Hengxin Tan[2*], Yi Liu[3,4*], Kuanglv Sun[5*], Ying Shi[1], Yuxin Zhai[1], Hao Lin[1], Guanghan Cao[3], Binghai Yan[2], Xianhui Chen[5], Tao Wu[5], Guang-Ming Zhang[1,6,†], Luyi Yang[1,6,†]

[1]*State Key Laboratory of Low Dimensional Quantum Physics, Department of Physics, Tsinghua University, Beijing 100084, China.*

[2]*Department of Condensed Matter Physics, Weizmann Institute of Science, Rehovot 7610001, Israel.*

[3]*School of Physics, Zhejiang University, Hangzhou 310058, China.*

[4]*Department of Applied Physics, Key Laboratory of Quantum Precision Measurement of Zhejiang Province, Zhejiang University of Technology, Hangzhou 310023, China.*

[5]*Hefei National Research Center for Physical Sciences at the Microscale, University of Science and Technology of China, Hefei, Anhui 230026, China*

[6]*Frontier Science Center for Quantum Information, Beijing 100084, China.*

[*] *These authors contributed equally to this work.*

[†]*E-mails: GMZ: gmzhang@mail.tsinghua.edu.cn; LY: luyi-yang@mail.tsinghua.edu.cn.*

**ABSTRACT**

**While correlated phenomena of flat bands have been extensively studied in twisted systems, the ordered states that emerge from interactions in the intrinsic flat bands of kagome lattice materials remain largely unexplored. The newly discovered kagome metal $CsCr_3Sb_5$ offers a unique and rich platform for this research, as its multi-orbital flat bands at the Fermi surface result in a complex interplay of pressurized superconductivity, antiferromagnetism, a structural phase transition, and density wave orders. Here, using ultrafast optical techniques, we provide strong spectroscopic evidence for a charge density wave transition in $CsCr_3Sb_5$, resolving previous ambiguities. Crucially, we identify rotational symmetry breaking that manifests as a three-state Potts-type nematicity. Our elastoresistance measurements directly demonstrate the electronic origin of this order, as the**

**rotational-symmetry-breaking $E_{2g}$ component of the elastoresistance shows a divergent behaviour around the transition temperature. This exotic nematicity results from the lifting of degeneracy of the multi-orbital flat bands, akin to phenomena seen in certain iron-based superconductors. Our study pioneers the investigation of ultrafast dynamics in flat-band systems at the Fermi surface, offering new insights into the interactions between multiple elementary excitations in strongly correlated systems.**

## INTRODUCTION

Kagome lattice provides a rich platform for exploring novel quantum states, emerging from the interplay between its frustrated corner-sharing triangular geometry and its intriguing electronic structure, which naturally features Dirac points, van Hove singularities, and flat bands (*1-4*). Recent studies have uncovered the topological iron- and cobalt-based kagome materials (*5, 6*), as well as distinctive properties potentially related to van Hove singularity instabilities in vanadium-based $AV_3Sb_5$ ($A$ = K, Rb, Cs) systems, including superconductivity (*7, 8*), unique charge orders (*9-12*), and nematicity (*13-16*). However, the flat bands, which greatly reduce bandwidth and amplify electron correlation effects (*17*), are located far from the Fermi surface in $AV_3Sb_5$ and other kagome materials. Consequently, their impact on transport properties is limited, motivating the search for intrinsic flat bands at the Fermi surface in kagome systems (*18*), which can give rise to complex low-energy excitations and the emergence of intertwined ordered states, such as magnetism, superconductivity, and nematicity (*19*).

Substituting vanadium with chromium atoms in $CsV_3Sb_5$ enables the adiabatic tuning of the multi-orbital flat bands to the Fermi level (*20*), while preserving the $D_{6h}$ kagome crystal structure (Fig. 1A) and band geometry (*21, 22*). Moreover, Cr atoms introduce magnetism and enhance electronic correlations (*23, 24*), leading to richer physical properties compared to its vanadium-based counterpart. $CsCr_3Sb_5$ thus becomes the first kagome material that simultaneously exhibits pressurized superconductivity ($T_c$ = 6.4 K under 4.2 GPa), density wave phases, and magnetism (*21*). Upon cooling, this material undergoes concurrent antiferromagnetic and a complex phase transitions at $T^*$ = 55 K, accompanied by a 1×4 periodic lattice distortion, which alters resistance and magnetic susceptibility at the ambient pressure (Figs. 1B–C) (*25*). However, such symmetry changes and charge density wave (CDW) have not been conclusively observed through angle-resolved photoemission spectroscopy (ARPES) and inelastic X-ray scattering measurements (*25-28*), leaving unresolved contradictions. The fundamental origin and nature of the phase transition and rotational symmetry breaking have been unclear.

Ultrafast optical techniques provide a powerful means to directly probe such symmetry changes and orders during complex phase transitions, with high sensitivity, femtosecond temporal resolution, and sub-micrometer spatial resolution (*29*). For instance, a CDW or nematicity can reduce higher rotational symmetries (such as $C_4$ or $C_6$) to $C_2$, which is detectable via reflectivity and birefringence measurements (*14, 30, 31*). The birefringence signal captures the difference in reflectivity along the two principal axes of the CDW or nematicity structures, with maximum and minimum values at ±45° relative to these axes (Supplementary Materials, Section 1). In time-resolved experiments, a pump pulse modulates the material's optical response through laser-induced excitation, and subsequent changes in quasiparticle states or order parameters are then tracked by a time-delayed probe pulse (Fig. 1D). Compared to static optical spectra, time-resolved techniques circumvent static background issues, enable the detection of ultrafast dynamics, and can distinguish multiple orders (*32-36*). However, despite the clear advantages, achieving the integration of low temperatures, strong magnetic fields, high spatial and temporal resolutions, and photon-energy tunability in a single ultrafast optical system remains technically challenging.

In this work, we systematically investigate the ultrafast dynamics of $CsCr_3Sb_5$ utilizing a powerful custom-built, multi-probe optical system (see Fig. 1D and Supplemental Materials, section S1) (*37, 38*). We first used time-resolved reflectivity (TRR) to identify the phonon bottleneck effect during the density wave gap opening around $T^*$ (Fig. 1E). Coherent phonons emerge and match the 1×4 CDW structure predicted by first-principle calculations. These features provide strong spectroscopic evidence for the CDW order below $T^*$. In contrast to $CsV_3Sb_5$, we found distinct multi-orbital nematicity, identified by anomalous anisotropic dynamics and elastoresistance measurements. This nematic order arises from the lifting of degeneracy between flat-band orbitals below $T^*$, which strongly resembles observations in certain iron-based superconductors (Fe-SCs) (*39-41*). Moreover, time-resolved birefringence (TRB) provides conclusive evidence of rotational symmetry breaking (Fig. 1F), allowing us to identify three-state Potts nematic domains, and directly monitor the intertwined order dynamics as they evolve with temperature. Our work not only clarifies multiple degrees of freedoms and their interplay in the correlated kagome metal $CsCr_3Sb_5$, but also paves the way for developing a universal microscopic theory and exploring potential applications of strongly electron correlation systems.

## RESULTS

### Charge density wave revealed by time-resolved reflectivity

Our temperature-dependent TRR measurements, shown in Figures 1E and 2A, reveal complex, non-monotonic quasiparticle relaxation dynamics closely related to a phase transition. Upon pump laser excitation, we observe an instantaneous change in transient reflectivity, followed by a double-exponential decay, comprising a fast and a slow component (Supplemental Materials, section S2). These characteristic timescales are around $\tau_{\mathrm{fast}}$ = 0.3 ps and $\tau_{\mathrm{slow}}$ = 6 ps at 3 K, which is comparable with previous reports in $CsV_3Sb_5$ (*16*). The slow component could involve interactions with the lattice, consistent with the enhanced electron-phonon coupling expected in an ordered CDW phase. Additionally, the transient dynamics remain unaffected by an out-of-plane magnetic field up to 6 T (Supplemental Materials, section S3). This implies that the antiferromagnetic order is already established below $T^*$, and the applied field is not strong enough to influence the dynamics.

Figures 2B–C display the extracted temperature-dependent amplitudes and lifetimes. As the system heats towards $T^*$, the amplitude of the slow component decreases dramatically, while its lifetime diverges, reminiscent of the phonon bottleneck effect during quasiparticle relaxation (*42*). These behaviors are consistent with the Rothwarf-Taylor model, commonly used to describe systems undergoing gap opening like in superconductors or density wave states (*43*). Since $CsCr_3Sb_5$ does not exhibit superconductivity under ambient pressure (*21*), the observed gap opening suggests the presence of density wave orders. The fitted zero-temperature gap values are $\Delta_0$ = 6.7 ± 2.0 meV and 3.3 ± 0.7 meV for the reconstructed principal axes $a_1$ and $a_2$, respectively (Supplemental Materials, section S5).

Meanwhile, our experiments reveal light-driven coherent phonon oscillations in the ordered phase, as shown in Fig. 2D, after subtracting the double-exponential decay background from the raw data and Fourier transforming. Probe polarization dependence measurements at 3 K reveal two prominent phonon frequencies at 0.66 and 0.88 THz (see Supplemental Materials, section S6 for the data along axis $a_1$), both showing two-fold rotational symmetry (Fig. 2E). The 0.66 THz mode exhibits a two-branch angular distribution, while the 0.88 THz mode features an asymmetric four-branch angular distribution. The observed frequencies and symmetries only agree with the 1×4 CDW structural distortion predicted by first-principles calculations (Supplemental Materials, sections S8 and S9), which is the most energy favorable configuration (*44*). As the temperature increases, the phonon modes soften, broaden, and eventually vanish at $T^*$ (Fig. 2F), further confirming their direct relation with the CDW transition (Supplemental Materials, section S7). However, despite our results present a consistent picture that the predominant order parameter is a non-magnetic 1×4 CDW, we cannot validate or expel the possibility of spin density wave or other possible wavevectors beyond our detection methods, which requires further experimental efforts to address these possibilities.

Although similar CDW-related phonon emergence and gap-opening behaviors have been observed in $CsV_3Sb_5$ (*16*), the behaviors in $CsCr_3Sb_5$ are distinct and strongly influenced by flat-band-amplified electron-electron correlations.

Previous first-principles calculations have shown phonon instability across nearly the entire Brillouin zone in non-magnetic structures (*23*), indicating strong electron-electron and electron-phonon interactions. Based on these calculations, the most likely magnetic ground state configuration is the swapped antiferromagnetic inverse Star-of-David (SA-ISD) phase (*23, 44*). Compared to the 2×2 antiferromagnetic inverse Star-of-David (ISD) phase, the magnetic moments of two pairs of next-nearest neighboring atoms in one 2×4 cell swap (Fig. 2G). This swapping lowers the total energy and breaks the six-fold rotational symmetry. Interestingly, the SA-ISD state consists of two 1×4 stripes connected by time-reversal gliding-mirror symmetry with the 2×4 spin-density-wave, which can explain the 1×4 structural modulations observed in previous X-ray diffraction measurements [31].

Moreover, the CDW gap in $CsCr_3Sb_5$ is approximately an order of magnitude smaller than in $CsV_3Sb_5$ (*16*). This unusually weak CDW gap ratio $\Delta_0/k_B T^* \approx 1$, is much smaller than the typical range of 2–9 observed in other well-known CDW materials, as seen in ultrafast dynamics studies (*16, 42, 45, 46*). This suggests that the phase transition in $CsCr_3Sb_5$ may be accompanied by additional coexisting orders apart from the 1×4 CDW order. Such orders are commonly found in strongly correlated systems but rarely seen in kagome materials. These additional orders are likely driven by multi-orbital electronic nematicity, as further supported by our subsequent optical anisotropy and elastoresistance measurements.

**Electron orbital nematicity induced by multi-orbital flat bands**

Figures 3A–B present a set of TRR signals at various probe polarization directions at 3 K, exhibiting significant anomalous anisotropy and a clear two-fold symmetry. In contrast, the TRR signals above the phase transition at 80 K (Fig. 3C) show no angular dependence and sign reversal behaviors. The extracted TRR amplitudes at 3 K (Fig. 3D) show opposite signs along the two principal axes (0° and 90°), which contrasts sharply with the near isotropic TRR signals in $CsV_3Sb_5$ (*47*) (Fig. 3E), suggesting a distinct order in the CDW state. This sign reversal in TRR is strikingly similar to results from some Fe-SCs (*39-41*), such as $NaFe_{1-x}Co_xAs$ (Fig. 3F), where electron orbital nematicity is the known cause.

The reflectivity is governed by the selection rules associated with the optical transition matrix for photon absorption, as well as the corresponding joint density-of-states (JDOS) linking the optical transition. Regarding the first factor, our work shows that the CDW gap is too small to significantly alter the overall band structure. In addition, three independent ARPES studies reveal negligible changes in the band structure in a large temperature range (*26-28*). As a result, the optical transition matrix for probe photon absorption is largely unaffected throughout the phase transition. Instead, the anisotropy arises from the second factor. In $CsCr_3Sb_5$, the flat bands near the Fermi level cause a divergence in the JDOS for transitions involving these bands, making them the dominant contributors to reflectivity change. In both $CsCr_3Sb_5$ and some Fe-SCs, the nematicity lifts the degeneracy of the multiple 3*d* orbitals around the Fermi level, which anisotropically distorts the Fermi surface. This results in an increase in the density of states (DOS) for some orbitals and a decrease for others, as evidenced by the first-principles calculations (Supplemental Materials, section S10). Consequently, upon cooling below the transition temperature, reflectivity increases for one polarization direction and decreases for the other, driven by polarization-dependent transitions involving distinct 3*d* orbitals. When the pump laser perturbs the ordered state through heating, the transient reflectivity shifts in the opposite direction, reflecting a reduction in the order parameter and giving rise to anomalous anisotropic TRR signals (Supplemental Materials, section S10).

We further confirm this nematicity with a clear elastoresistance response near the transition temperature. Using the modified Montgomery technique (Supplemental Materials, section S15), we measured the rotational-symmetry-breaking $E_{2g}$ component of the elastoresistance as a function of temperature (Fig. 3G). Critically, this component shows a divergent behavior around $T^*$, which is a hallmark of the electronic origin of nematicity (*48*). This stands in stark contrast to the abrupt "step jump" around the CDW transition temperature seen in $AV_3Sb_5$ (*13, 49-51*), indicating the absence of electronic nematic instability. Above the transition temperature, the anisotropic $E_{2g}$ component follows a Curie-Weiss behavior, indicating the gradual establishment of the nematic order. This peak of nematic fluctuations around the transition temperature is consistent with Landau's phase transition theory. Combining with our calculated orbital splitting, these results suggest that orbital reconstruction is the driving mechanism for nematicity rather than a mere consequence of the structural transition.

The elastoresistance data further strengthen the analogy to some Fe-SCs, where similar electronically-driven anisotropic responses were found in $Ba(Fe_{1-x}Co_x)_2As_2$ (*48, 52*) and concurrent density wave and nematic orders phase transition were reported in $BaNi_2As_2$ (*53*). The peak $E_{2g}$ elastoresistance value of ~20 is moderate and comparable to values reported in $BaNi_2(As_{1-x}P_x)_2$ (*54*) and $Ba(Fe_{1-x}Co_x)_2As_2$ (*55*). In addition, from the perspective of symmetry, it is worth noting that nematicity in $CsCr_3Sb_5$ is of the Potts type ($C_6$ to $C_2$), while in some Fe-SCs, it is the Ising type ($C_4$ to $C_2$). The presence of nematicity in $CsCr_3Sb_5$ is also supported by TRB data, as discussed in the following section.

**Rotational symmetry breaking and three-state Potts nematicity**

To gain deeper insights into the CDW and nematicity in $CsCr_3Sb_5$, we conducted TRB measurements, which are sensitive to rotational symmetry breaking and can directly capture the dynamics of the emerging orders. Figure 4A shows a set of TRB signals at various probe polarization directions after laser excitation at 3 K. The TRB signal amplitude modulates as the polarization direction rotates, indicating $C_6$ rotational symmetry breaking. Moreover, a real-space scan with a 2 μm diameter light spot reveals three different domains, each with its principal axis rotated by 120° relative to the others (Fig. 4B), clearly confirming a three-state Potts nematicity, similar to that observed in $CsV_3Sb_5$ (*14*). However, the typical domain size in $CsCr_3Sb_5$ is estimated to be around 10 μm at 3 K, which is approximately an order of magnitude smaller. This pattern disappears above the transition temperature $T^*$ (e.g. at 60 K in Fig. 1F), indicating the loss of nematic order and the recovery of six-fold rotational symmetry. Additionally, the TRB dynamics are unaffected by an out-of-plane magnetic field up to 6 T (Supplemental Materials, section S3), consistent with TRR measurements.

Figure 4C displays the temperature-dependent TRB signals at a fixed probe angle of 45° relative to the principal axes within a given domain, which exhibit a complex evolution with temperature. We analyzed the signals using a double-exponential fit (Supplemental Materials, section S12) and plotted the fitting parameters as a function of temperature in Figs. 4D–E. The two decay processes show notable differences: the lifetime of the slow decay process $\tau_{\text{slow}}$ diverges around $T^*$ (Fig. 4E), resembling the same phonon bottleneck effect observed in TRR, while the $\tau_{\text{fast}}$ maintains a stable sub-picosecond lifetime around $T^*$, consistent with the earlier discussions on the electron orbital nematic order. The presence of coexisting, intertwined orders highlights the unusual and complex nature of the phase transition in $CsCr_3Sb_5$, and distinguishes it from $AV_3Sb_5$ (which has a 2×2×2 CDW) and $ATi_3Bi_5$ (which lacks a CDW) (*56*).

## DISCUSSION

Figure 4F summarizes the complex correlated physics in $CsCr_3Sb_5$ by integrating our findings on charge and orbital orders with existing literature on lattice and spin degrees of freedom. These degrees of freedom interact on the platform of flat bands at the Fermi surface. Importantly, the extremely strong electron correlations in the flat bands can induce significant spin fluctuations (*21-24, 26-29*), which are further enhanced by the geometric frustration of the kagome lattice, evidenced by a Curie–Weiss temperature exceeding 300 K [25]. We propose that lattice distortions arise to compensate and balance the strong spin fluctuations, leading to the formation of the electron order. Moreover, the orbital splitting reflects a combination of both structural and orbital orders, with the latter contributing to the further amplification of nematicity rather than by a pure structural transition. These complex interactions result in the surprising emergence of multiple long-range orders within a narrow temperature range. Our overview not only highlights the interactions among degrees of freedom and reveals strong correlations within multi-orbital flat bands, but also paves the way for constructing a universal microscopic theory.

Last but not least, our observations of gap opening, crystal reconstruction, and rotational symmetry breaking associated with the phase transition in $CsCr_3Sb_5$, contrast with previous ARPES studies, which reported nearly invariant band structures, $C_6$-symmetrized constant-energy contours and undetectable CDW (*26-28*). These apparent discrepancies can be attributed to two main factors. First, the CDW gap is quite small, falling below the energy resolution of standard ARPES techniques. Second, the difference in spatial resolutions between our setup (~2 μm) and typical ARPES experiments (> 30 μm) is crucial. ARPES may average out signals from different domains, thereby masking rotational symmetry breaking and CDW features that we directly observe.

In conclusion, by utilizing multi-probe ultrafast optical techniques and elastoresistance measurements, we have provided a comprehensive picture of the complex correlated phase transition in the kagome metal $CsCr_3Sb_5$. We provide strong spectroscopic evidence for a small-gap nematic 1×4 CDW and uncover an unexpected electron orbital nematic order driven by strong electron correlations, resembling orbital nematicity observed in some Fe-SCs.

Additionally, we have revealed a three-state Potts nematicity in the ordered phase. By demonstrating the intricate interactions among spin, charge, orbital, and lattice degrees of freedom, our work not only advances the research on this newly synthesized strongly correlated kagome metal, but also deepens the understanding of the rich physics associated with intrinsic flat bands.

## MATERIALS AND METHODS

### Sample growth and characterization

Single crystals of $CsCr_3Sb_5$ flakes with a typical size of 0.5 × 0.5 × 0.02 $mm^3$ were grown via the self-flux method (*21*). Crystals were subsequently characterized by X-ray diffraction and energy-dispersive X-ray spectroscopy. Magnetic measurements were conducted using a Magnetic Property Measurement System (MPMS-3, Quantum Design). Resistivity measurements were performed using the standard four-terminal method.

### Cooperative multi-probe ultrafast optical measurements

The sample was cleaved and kept in the vacuum chamber of an optical superconducting magnet system during the experiments. We conducted non-degenerate two-color pump-probe experiments using a Ti:sapphire oscillator paired with an optical parametric oscillator (OPO), operating at an 80 MHz repetition rate. The pump beam was centered at 610 nm, while the probe beam wavelength varied from 740 nm to 860 nm, primarily centered around 780 nm. Both beams were collinearly focused to 3 μm (pump) and 2 μm (probe), respectively, using a customized non-magnetic low-temperature objective (40×, NA = 0.5). The overall temporal resolution of the setup was 250 fs. To improve the signal-to-noise ratio, the intensity of the pump beam was modulated by an Electro-Optic Modulator (EOM) at 473 kHz to enable lock-in detection. The reflected probe light passed through a long-pass filter to remove residual pump light before reaching the detector. In the TRB and TRMOKE experiments, the detection of the optical polarization rotation was achieved with a standard optical bridge arrangement using balanced photodiodes, which effectively mitigated laser power fluctuations. In the TRR measurements, we monitored the pump-induced change in reflectivity by using a photodiode to measure the probe reflection. See Supplementary Materials Section 1 for further details.

### First-principles calculations

Density functional theory (DFT) calculations were conducted using the Vienna Ab-initio Simulation Package (VASP) (*57*). The exchange-correlation interactions were treated using the generalized gradient approximation (GGA) as parametrized by Perdew-Burke-Ernzerhof (PBE) (*58*). A plane-wave energy cutoff of 300 eV was applied throughout the calculations. The pristine structure was fully relaxed in a ferromagnetic configuration until the residual forces on the atoms were reduced to below 5 meV/Å. A k-point grid of 12×12×6 was used to sample the Brillouin zone for the pristine phase. For the CDW phase, the symmetry of the altermagnetic configuration (*23*), was initially used to induce artificial distortion in the superstructure. The distorted structure was then fully relaxed with a force threshold of 5 meV/Å. Subsequently, the phonon spectrum was computed using the finite-displacement method, as implemented in the Phonopy software (*59*). Spin-orbital coupling is omitted in all calculations.

## Elastroresistance Measurements

The elastoresistance measurements were conducted in a modified Montgomery configuration. In this configuration, a square-shaped sample (0.29 mm × 0.30 mm × 0.02 mm) was glued onto a piezoelectric stack (Piezomechanik PSt150/5×5/7), and electrical contacts were made at four corners. Uniaxial strain was applied by applying a voltage to the PZT stack via a Keithley 2400 source meter. This voltage induced an expansion along the $x$ direction ($\varepsilon_{xx}$) and a corresponding Poisson contraction along the $y$ direction ($\varepsilon_{yy}$). The resultant strain was directly monitored by a strain gauge (WK-05-062TT-350, Vishay Precision Group) mounted on the PZT stack with high-performance epoxy (M-Bond 43-B, Vishay Precision Group). For resistance measurements, $R_{xx}$ and $R_{yy}$ were determined by applying current between contacts on one side of the sample and measuring the voltage on the opposite side via a Keithley 2182 nanovoltmeter and a Keithley 6221 current source. The measured resistances were then converted to the corresponding resistivities $\rho_{xx}$ and $\rho_{yy}$ following the established geometric conversion procedure in Refs. 48 and 49. By combining the elastoresistance coefficient in the $x$- and $y$-directions, we can obtain the elastoresistance responses for $E_{2g}$ symmetry. The formula of the elastoresistance response for $E_{2g}$ symmetry is given as:

$$m_{E_{2g}} = \frac{1}{(1+v_{xy})}\left(\frac{d\left(\frac{\rho_{xx}}{\rho_0}\right)}{d\varepsilon_{xx}} - \frac{d\left(\frac{\rho_{yy}}{\rho_0}\right)}{d\varepsilon_{xx}}\right),$$

where $v_{xy} = -\varepsilon_{yy}/\varepsilon_{xx}$ represents the Poisson's ratio of the PZT stack.

## Supplementary Materials

This PDF file includes:

Supplementary text

Figures S1 to S12

Table S1

**Acknowledgements**

We thank Shiliang Li, Bo Liu, Zhiyuan Sun, F. Michael Bartram, and Qiong Wu for their helpful discussions. **Funding:** The work was supported by the National Natural Science Foundation of China (Grants No. 12361141826, No. 12421004, and No. 12074212), the National Key R&D Program of China (Grants No. 2021YFA1400100 and No. 2020YFA0308800), and the Beijing Natural Science Foundation (Grant No. Z240006). B.Y. acknowledges the financial support by the Israel Science Foundation (ISF: 2932/21, 2974/23), German Research Foundation (DFG, CRC-183, A02), and by a research grant from the Estate of Gerald Alexander. G.M.Z. acknowledges the support of the National Key Research and Development Program of China (Grants No. 2023YFA1406400). Y.L. acknowledges the support of the National Natural Science Foundation of China (Grants No. 1247042667). **Author contributions:** L.Y. conceived the projects. L.Y.L. and Y.D.L. carried out time-resolved optical measurements, Raman experiments, and corresponding data analysis with the help of Y.X.Z., Y.S., H.L., and L.Y. *Ab-initio* calculations were performed by H.X.T. and B.H.Y. Single crystals were synthesized and characterized by Y.L. and G.H.C. Elastoresistance measurements were carried out by K.L.S., T.W., and X.H.C. G.M.Z. proposed the physical picture to understand the experimental results. The paper was written by L.Y.L., Y.D.L., and L.Y. with the help of G.M.Z. All authors contributed to the scientific planning and discussions. **Competing interests:** The authors declare that they have no competing interests. **Data and materials availability:** The data sets that support the findings of this study are available from the corresponding author upon reasonable request.

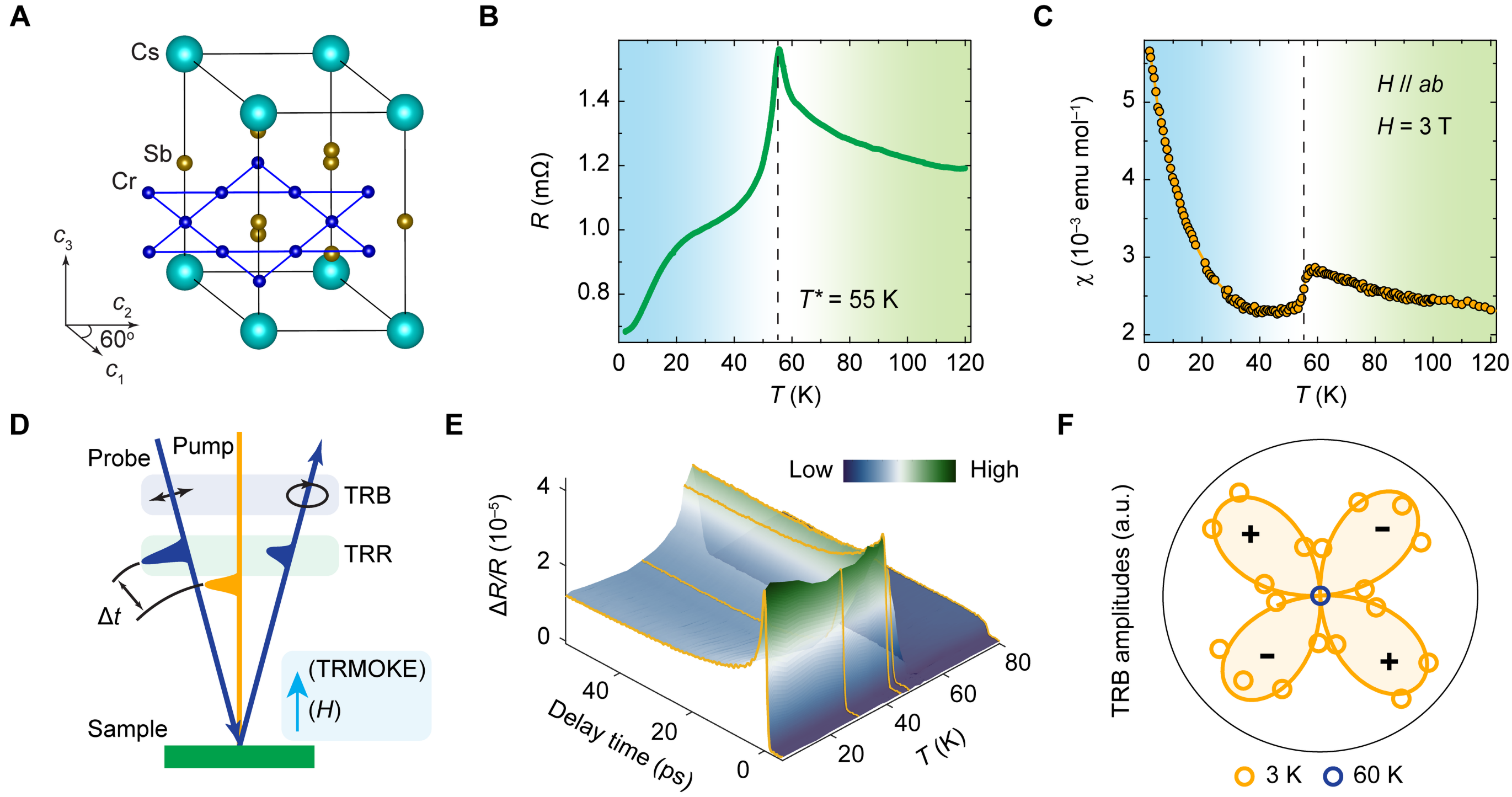

**Fig. 1. Basic properties of correlated kagome metal $CsCr_3Sb_5$.** (**A**) Crystal structure of $CsCr_3Sb_5$ in the normal phase, where Cr atoms form kagome sheets. (**B–C**) Temperature dependence of resistance (B) and magnetic susceptibility with an in-plane magnetic field $H$ = 3 T (C) both indicating a phase transition at $T^*$ = 55 K. (**D**) Schematic illustration of the multi-probe ultrafast optical system. TRB: time-resolved birefringence; TRR: time-resolved reflectivity; TRMOKE: time-resolved magneto-optical Kerr effect. (**E**) Three-dimensional plot of the temperature-dependent TRR signals across the phase transition. (**F**) Angular distribution pattern of TRB signal amplitudes, indicating rotational symmetry breaking with multiple coexisting orders throughout the phase transition.

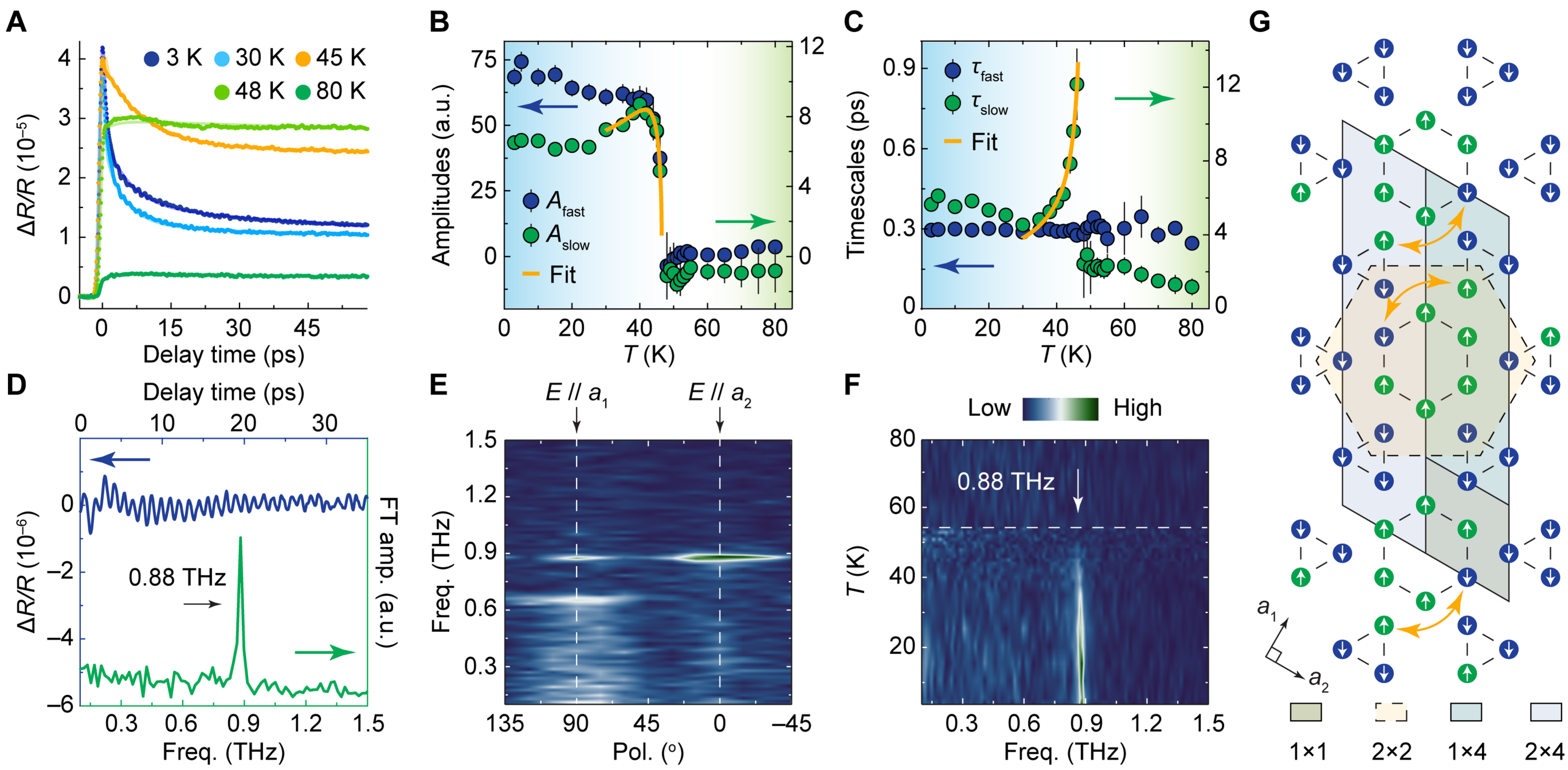

**Fig. 2. 1×4 charge-density-wave (CDW) state in $CsCr_3Sb_5$ detected by TRR.** (**A**) TRR signals and double-exponential fittings of orange lines in Fig. 1E. (**B–C**) Extracted temperature-dependent amplitudes (B) and timescales (C) of TRR signals in Fig. 1E. (**D**) Residual oscillatory component of the signal (blue) after subtracting the double-exponential background and corresponding Fourier transform (FT) amplitude (green) at $T = 3$ K. (**E**) FT amplitudes as a function of probe polarization directions (E) at $T = 3$ K. (**F**) Temperature-dependent FT intensities map, with the white dashed line marking $T^*$. (**G**) Schematic illustration of the swapped antiferromagnetic inverse Star-of-David (SA-ISD) phase, featuring a 1×4 CDW and a 2×4 spin-density wave (SDW) modulations with broken rotational symmetry. The diagram also shows the 1×1 normal and the 2×2 inverse Star-of-David (ISD) structures. All data were collected under the pump fluence of $F = 200$ μJ cm$^{-2}$, with the pump wavelength of 610 nm, and the probe wavelength of 780 nm. Data in panels A–C were collected along the principal axis $a_1$, while in panels d and f along $a_2$. Data along the other axes are available in the Supplemental Materials.

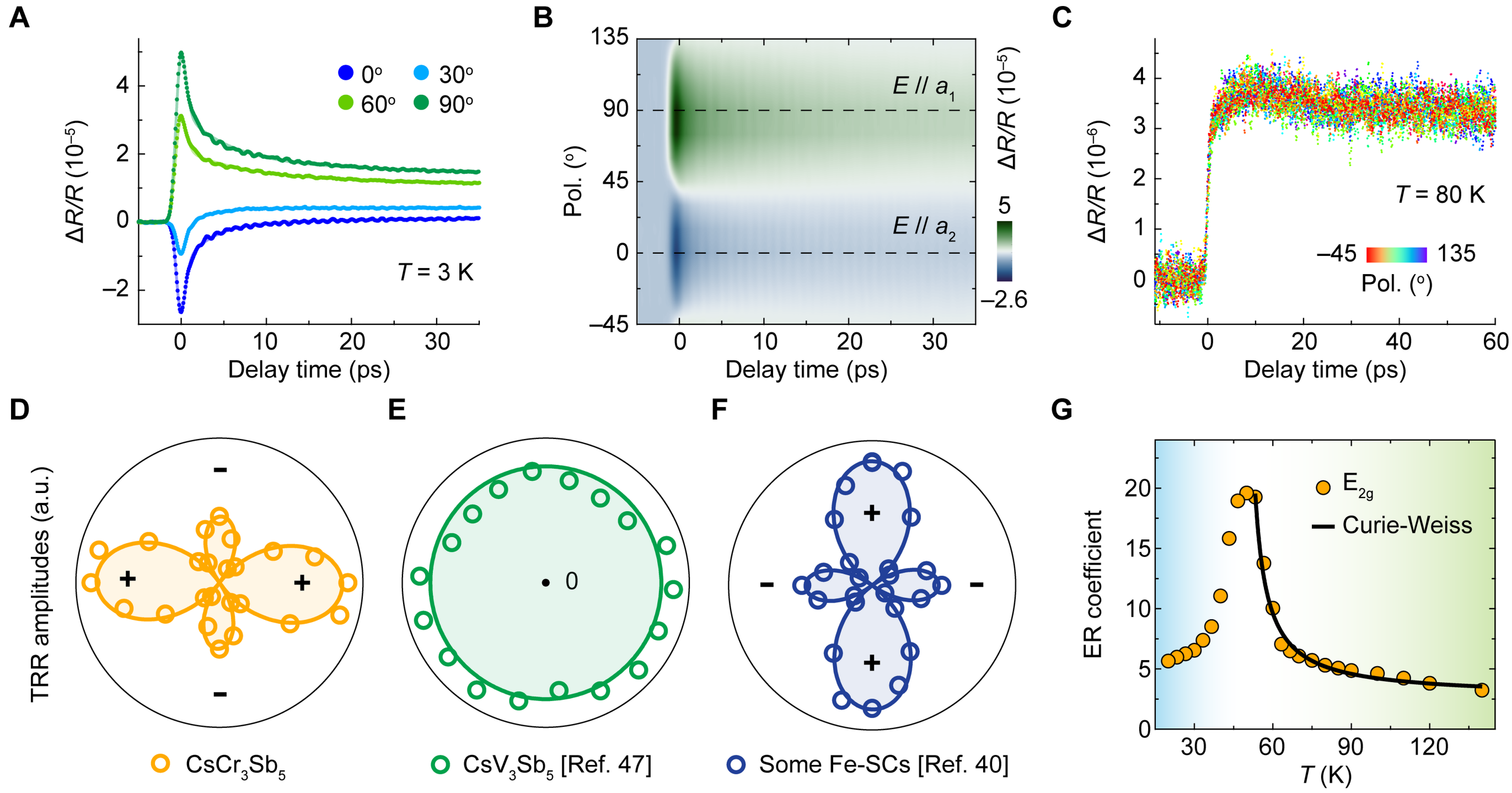

**Fig. 3. Electron orbital nematicity of $CsCr_3Sb_5$.** (**A**) Anisotropic TRR signals at selected probe polarization directions. (**B**) TRR signals as a function of probe polarization directions. (**C**) Isotropic TRR behavior above $T^*$. (**D**–**F**) Angular distribution patterns of TRR amplitudes. $CsCr_3Sb_5$ and some iron-based superconductors [Fe-SCs, e.g. $NaFe_{1-x}Co_xAs$ (*40*)] display a similar asymmetric 4-branch pattern, in contrast to an almost circular pattern of $CsV_3Sb_5$ (*47*). (**G**) Temperature-dependent anisotropic $E_{2g}$ component of the elastoresistance. The black line represents the Curie-Weiss fitting to the data above $T^*$. All data were collected under the pump fluence of $F = 200$ μJ cm$^{-2}$, wavelength of 610 nm, and the probe wavelength of 780 nm. Data were collected at $T = 3$ K in panels A and B, and $T = 80$ K in panel C.

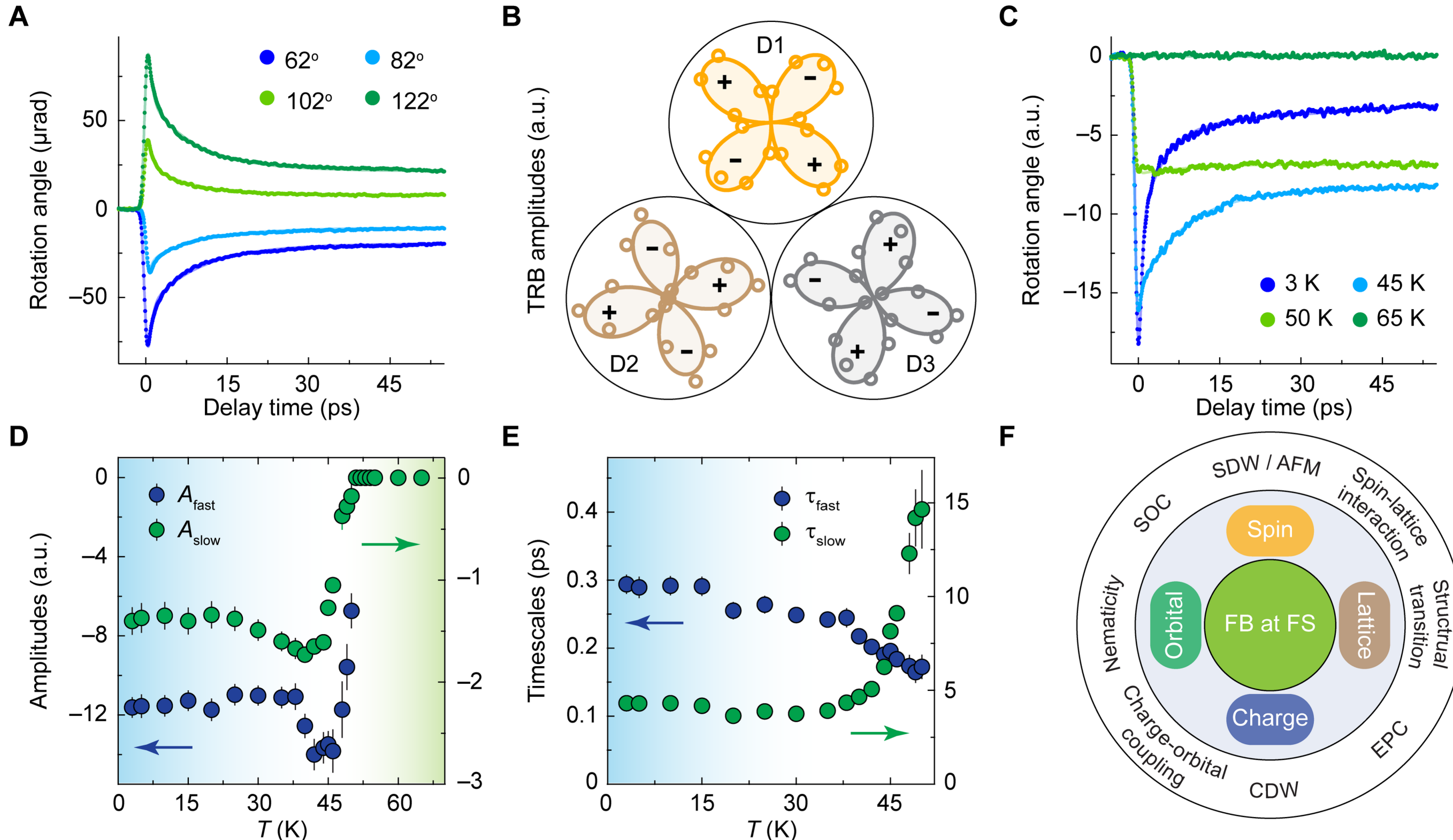

**Fig. 4. Three-state Potts nematicity and intertwined orders in $CsCr_3Sb_5$.** (**A**) TRB signals for selected probe polarization directions at $T = 3$ K. (**B**) Angular distribution patterns of TRB signals in three adjacent domains. (**C**) Temperature-dependent TRB signals within a single domain. The probe polarization direction is set at 45° relative to the principal axes. (**D–E**) Temperature-dependent amplitudes (D) and timescales (E) of TRB signals, extracted from double-exponential fitting. (**F**) Schematic illustration of the interplay between spin, orbital, lattice, and charge degrees of freedom, set on the platform of flat bands (FB) at the Fermi surface (FS), with their intrinsic properties and interactions. AFM: antiferromagnetism; SOC: spin-orbital coupling; EPC: electron-phonon coupling. All data were collected under the pump fluence of $F = 200$ μJ cm$^{-2}$, with the pump wavelength of 610 nm, and the probe wavelength of 780 nm.